\documentstyle[epsfig]{mn2e}

\def\gtsima{$\; \buildrel > \over \sim \;$}
\def\ltsima{$\; \buildrel < \over \sim \;$}
\def\gtrsim{\lower.5ex\hbox{\gtsima}}
\def\lesssim{\lower.5ex\hbox{\ltsima}}

\begin{document}

\title[Ring galaxies from off-centre collisions]{Ring galaxies from off-centre collisions}
\author[Mapelli \&{} Mayer]
{M. Mapelli$^{1}$ \&{} L. Mayer$^{2}$
\\
$^1$INAF-Osservatorio astronomico di Padova, Vicolo dell'Osservatorio 5, I--35122, Padova, Italy, {\tt michela.mapelli@oapd.inaf.it}\\
$^2$Institute for Theoretical Physics, University of Z\"urich, Winterthurerstrasse 190, CH--8057, Z\"urich, Switzerland\\
}
\maketitle \vspace {7cm }

\begin{abstract}
We investigate the formation of RE galaxies (i.e. of collisional ring galaxies with an empty ring), with N-body/SPH simulations.
 The simulations 
employ a recipe for star formation (SF) and feedback
that has been shown to be crucial to produce realistic galaxies in a cosmological context.
 We show that RE galaxies can form via off-centre collisions (i.e. with a non-zero impact parameter), even for small inclination angles. The ring can be either a complete ring or an arc, depending on the initial conditions (especially on the impact parameter). In our simulations, the nucleus of the target galaxy is displaced from the dynamical centre of the galaxy and is buried within the ring, as a consequence of the off-centre collision. We find that the nucleus is not vertically displaced from the plane of the ring. We study the kinematics of the ring, finding agreement with the predictions by the analytic theory. The SF history of the simulated galaxies indicates that the interaction enhances the SF rate. We compare the results of our simulations with the observations of Arp~147, that is the prototype of RE galaxies.
\end{abstract}

%

\section{Introduction}
Ring galaxies are characterized by a bright ring of gas and stars, whose diameter may be very large ($\le{}100$ kpc, Ghosh \&{} Mapelli 2008) and where star formation (SF) is generally very intense (Higdon 1995, 1996; Marston \&{} Appleton 1995; Mayya et al. 2005; Romano, Mayya \&{} Vorobyov 2008). 
A large fraction of ring galaxies ($\approx{}60$ per cent, Few \&{} Madore 1986) 
is thought to form via (almost) head-on collisions with massive intruder 
galaxies (at least one tenth of the target mass). Because of the gravitational perturbation induced by the bullet 
galaxy, a density wave propagates through the disc of the target galaxy, 
generating an expanding ring of gas and stars (e.g., Lynds \&{} Toomre
1976; Theys \& Spiegel 1976; Toomre 1978; Appleton \&{} James 1990; Appleton \&{} Struck-Marcell 1987a, 1987b; Struck-Marcell \&{} Lotan 1990; Hernquist \& Weil 1993; Mihos \& Hernquist 1994;  Appleton \& Struck-Marcell 1996; Gerber, Lamb \&{} Balsara 1996; Struck 1997; Horellou \& Combes 2001; Mapelli et al. 2008a, 2008b). Therefore, ring galaxies are unique laboratories where the effects of galaxy interactions can be studied, from a plethora of
different points of view. First, the kinematics of the ring provides
information about the dynamics of the interaction. The simple geometry
of ring galaxies, combined with the essential kinematic data, allows
us to reconstruct the main features of the interaction and the time
elapsed since it occurred. Furthermore, ring galaxies are often
characterized by high SF rate (SFR, up to $\approx{}20$
M$_\odot{}$ yr$^{-1}$), suggesting that the density wave associated
with the propagating ring triggers the formation of stars. When the
density wave reaches the external region of the target galaxy, SF may involve `fresh' gas, almost unpolluted by previous
episodes of SF.  This likely explains
why the only three ring galaxies (the Cartwheel, AM1159-530 and Arp 284) for which metallicity measurements
have been published (Fosbury \&{} Hawarden 1977; Weilbacher, Duc \&{} Fritze-v. Alvensleben  2003; Smith, Struck \&{} Nowak 2005) have metallicity Z $\le0.3$ Z$_\odot{}$ (when recalibrated on the basis of the electron temperature or of the P-method, see, e.g., Pilyugin, V\'ilchez \& Thuan 2010).

One particular subclass of ring galaxies, named RE by Theys \& Spiegel (1976), is characterized by the fact that the ring is empty at its interior, apparently lacking the nucleus. This class includes objects that are among the closest and the most studied ring galaxies (e.g., Arp 146, Arp 147, VII Zw 466, I Zw 28, NGC~985, AM~0058-220, AM~2145-543). Arp 147 is generally considered the prototype of this class. Arp~147 has a high SFR ($\sim{}4-12$ M$_\odot{}$ yr$^{-1}$, Romano et al. 2008; Rappaport et al. 2010; Fogarty et al. 2011), and a relatively young ring: $\approx{}50$ Myr elapsed after the collision, as estimated on the basis of the ring expansion velocity ($113\pm{}4$ km s$^{-1}$, Fogarty et al. 2011) and diameter ($\sim{}15$ kpc, de Vaucouleurs et al. 1991).  Arp 147 has a total stellar mass similar to the Milky Way ($6-10\times{}10^{10}$ M$_\odot{}$, Fogarty et al. 2011). A fraction $\sim{}0.04-0.34$ of the total stellar light is contributed by a young ($1-70$ Myr) stellar population (Fogarty et al. 2011). Recently, Rappaport et al. (2010) reported the detection of 9 ultra-luminous X-ray sources (ULXs, i.e. non-nuclear point-like X-ray sources with luminosity $L_X\gtrsim{}10^{39}$ erg s$^{-1}$) in the ring of Arp 147.

Lynds \&{} Toomre (1976) proposed that an off-centre collision can displace the nucleus and make it appear embedded in the ring. This idea is supported by observations (e.g., Thompson \&{} Theys 1978; Schultz et al. 1991), that show that the suspected remnant nucleus can be distinguished from the other knots of the ring for its lack of SF.

Gerber, Lamb \&{} Balsara (1992) confirm, by running N-body/smoothed particle hydrodynamics (SPH) simulations, that RE galaxies form via off-centre collisions. They propose that the ring of RE galaxies is incomplete and may appear a complete ring because of the viewing angle. They find that the remnant nucleus is not only off-set from the centre, but also out of the plane of the ring, and appears to be buried within the ring only because of projection effects. On the other hand, Fogarty et al. (2011) find no evidence for the offset of the nucleus from the plane of the disc, based on kinematics.

In this paper, we run N-body/SPH simulations of off-centre galaxy collisions, to study the formation of RE galaxies. The simulations are described in Section 2. Section 3 reports the results and Section 4 summarizes the main conclusions.
In our simulations, we adopt initial parameters close to the properties of Arp~147. However, we are not interested in producing a `perfect' model of Arp~147, but in understanding the formation of RE galaxies, from a more general point of view.

\section{Simulations}
The initial conditions for both the target and the intruder galaxy are generated by using an upgraded version of the code described in Widrow, Pym \&{} Dubinski (2008; see also Kuijken \&{} Dubinski 1995 and Widrow \&{} Dubinski 2005). The code generates self-consistent disc-bulge-halo galaxy models, derived from explicit distribution functions for each component, that are very close to equilibrium. In particular, the halo is modelled as a Navarro, Frenk \&{} White (1996, NFW) profile. The disc (when present) is exponential. The code also includes the possibility of adding a Hernquist bulge (Hernquist 1993). We generate bulgeless models for the target galaxy and models with  bulge for the bullet galaxy.

Masses and characteristics lengths of the simulated galaxies are listed in Tables~1 and 2. We report the results of eight runs. 
Runs~A, B, C and D will be considered our fiducial models, whereas runs~E, F, G and H were performed for comparison with Gerber et al. (1992).
The only differences among runs A, B, C and D are the initial mass of gas and stars of the target galaxy. The total initial mass in baryons (gas and stars) is the same ($8.48\times{}10^{10}$ M$_\odot{}$) for all the simulations, but the ratio between gas and star mass ($f{\rm gas}$) is different: $f_{\rm gas}=0.5$, $0.2$, $0.09$ and $0.05$, in run~A, B, C and D, respectively.  These values of $f_{\rm gas}$ cover the typical range observed for Milky Way-like galaxies (see, e.g., Haynes et al. 1999; Geha et al. 2006).

Runs~E, F, G and H have the same star and gas content as run~D. They differ from run~D for the impact parameter (runs~E and G) and/or for the target-to-companion mass ratio (runs~F, G and H). Run~H differs from the other runs also for the halo-to-baryon mass ratio ($\sim{}2.5$ instead of $\sim{}8$, for comparison with Gerber et al. 1992) and for the zero inclination angle.

We set the mass of the companion to be approximately half of (equal to) the mass of the target in runs A, B, C, D and E (F, G and H).\footnote{Romano et al. (2008) report a mass ratio $M_{\rm targ}/M_{\rm int}=$0.57 between the ring galaxy and the intruder in the Arp~147 system. They derive the two masses from the K magnitude and from the colours (using the recipes by Bell et al. 2003). According to their estimates the stellar mass of the ring galaxy Arp~147 is $M_{\ast}=2.11\times{}10^{10} M_\odot{}$. On the other hand, Fogarty et al. (2011) estimate a stellar mass of the ring galaxy Arp~147 $M_{\ast}=6.1-10.1\times{}10^{10}$ M$_\odot{}$, by combining optical measurements and galaxy population models. Since Bell et al. (2003) models are known to be less accurate in the case of late-type galaxies than in the case of early-type galaxies, it is reasonable to adopt the mass estimate by Romano et al. (2008) for the intruder galaxy and that by Fogarty et al. (2011) for the target. Therefore, the mass ratio between target and intruder is $M_{\rm targ}/M_{\rm int}=1.6-2.7$ (assuming that the dynamical mass scales approximately as the stellar mass). Consequently, we adopt $M_{\rm targ}/M_{\rm int}=2$ in runs A, B, C, D and E. We also make some check runs (F, G and H), where the mass of the target is the same as the mass of the intruder, for comparison with Gerber et al. (1992).}

We initially set the centres of the target and the bullet galaxy at a distance of three virial radii of the target. For runs A, B, C, D, F and H, the impact parameter is $b=8$ kpc. In runs E and G $b$ is $=12$ kpc. 
In all runs, the initial relative velocity between the centres of mass is $v_{\rm rel}=900$ km s$^{-1}$ (see Mapelli et al. 2008a). 
The inclination angle (with respect to the symmetry axis of the target) is $\alpha{}=7^{\circ{}}$, with the exception of run~H (where $\alpha{}=0$ for comparison with  Gerber et al. 1992).
 The inclination angle and the relative velocity are slightly different from those obtained for Arp~147 by the most recent measurements\footnote{Fogarty et al. (2011) constrain the line-of-sight maximum distance between the two galaxies to be $l=7.8$ kpc and measure a projected distance of $d=12.8$ kpc. They also estimate the empty ring to be inclined by $i=25^\circ$ with respect to the plane of the sky. From simple geometrical arguments, this leads to a minimum inclination angle for the interaction of $\approx{}33^\circ$, larger than our fiducial model (in our model, $\alpha{}=7^{\circ{}}$ and $v_{\rm rel}=900$ km s$^{-1}$ imply $l=30$ kpc, $d=15$ kpc and  $i=20^\circ$).}. In the Appendix~A, we show a simulation performed with $\alpha{}=33^\circ$, in agreement with the results by Fogarty et al. (2011). However, the morphology of the ring formed in this simulation looks very different from that of Arp~147, as too much warping is induced by such a large inclination angle. We will investigate the role of inclination angle and warping in a forthcoming paper. Furthermore, we stress again that the aim of this paper is to investigate the formation of RE galaxies, rather than producing a perfect model of Arp~147.


\begin{table}
\begin{center}
\caption{Initial conditions.}
 \leavevmode
\begin{tabular}[!h]{lll}
\hline
& Target & Bullet \\
\hline
M$_\ast{}$$^{\rm a}$ [$10^{10}$ M$_\odot{}$]      & see Table~2 & 1.91\\
Gas Mass$^{\rm b}$ [$10^{10}$ M$_\odot{}$]    & see Table~2 & $-$ \\
Halo scale length$^{\rm c}$ [kpc] & 6.0 & 6.0\\
Disc scale length [kpc] & 3.7 & 3.0 \\
Disc scale height [kpc] & 0.37 & 0.30 \\
Bulge scale length [kpc] & $-$ & 0.6 \\
\noalign{\vspace{0.1cm}}
\hline
\end{tabular}
\footnotesize{\\$^{\rm a}$  M$_\ast{}$ is the total stellar mass of the galaxy. The stars of the target are distributed according to an exponential disc. The stars of the bullet are distributed according to an exponential disc (75 per cent of the total stellar mass) and an Hernquist bulge (25 per cent).$^{\rm b}$ The gas of the target is distributed according to an exponential disc, with the same parameters (scale length and height) as the stellar disc. $^{\rm c}$ We name halo scale length the NFW scale radius $R_{\rm s}\equiv{}R_{200}/c$, where $R_{200}$ is the virial radius of the halo (NFW 1996) and $c$ the concentration (here we assume $c=12$ for both galaxies).}
\end{center}
\end{table}

The mass resolution of the simulations is $10^5$ M$_\odot{}$ for stars and gas, and $5\times{}10^5$ M$_\odot{}$ for dark matter (DM). The softening length is 0.1 kpc.
We simulate the evolution of the models with the N-body/SPH tree code gasoline (Wadsley, Quinn \&{} Stadel 2004). Radiative cooling, SF
 and supernova (SN) blastwave feedback are enabled, as described in Stinson et al. (2006, see also Katz 1992). The adopted parameters for SF and feedback (see Section~3.4) are the same as used in recent cosmological simulations capable of forming realistic galaxies in a wide range of masses (e.g., Governato et al. 2010; Guedes et al. 2011).

\begin{figure}
\center{{
\epsfig{figure=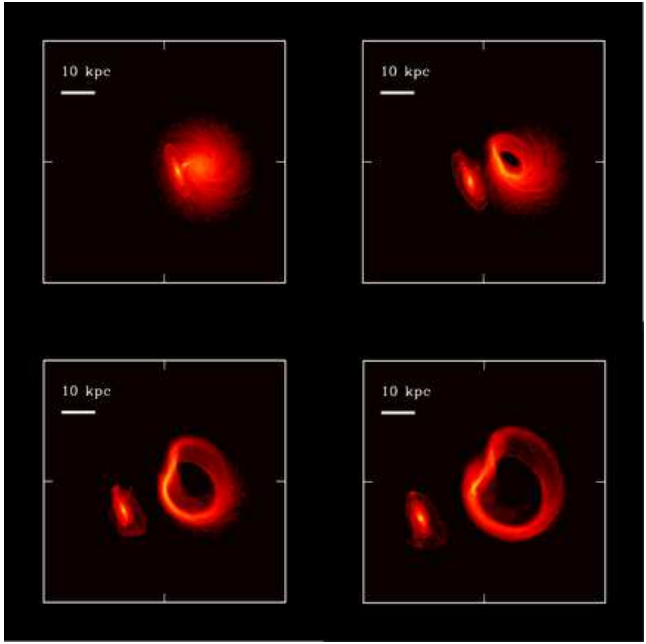,width=8.5cm} 
}}
\caption{\label{fig:fig1}
Projected mass density of stars and gas in run C. The simulated target galaxy has been rotated by 20$^{\circ}$ along the $x$-axis in the direction of the $y$-axis and by 20$^{\circ}$ counterclockwise, to match the observations of Arp~147. The scale is logarithmic, ranging from $11.15$ to $7.04\times{}10^4$ M$_\odot{}$ pc$^{-2}$. From top to bottom and from left to right: 10, 30, 50 and 70 Myr after the interaction.
}
\end{figure}

\section{Results}
\subsection{Morphology of the ring in runs A, B, C and D}

We first consider the morphological evolution of the ring in our four fiducial runs (A, B, C and D).
Fig.~\ref{fig:fig1} shows the time evolution of run~C. The ring starts developing $20-30$ Myr after the collision. About 50 Myr after the  collision the diameter of the ring is $15-19$ kpc (similar to the observed diameter of Arp~147). We notice that the target galaxy develops a complete ring. 

Another interesting feature of Fig.~\ref{fig:fig1} is the structure of the companion galaxy. We simulate the bullet as an early-type spiral galaxy (with a purely stellar disc and without gas). The initial conditions of the bullet were chosen to allow for the formation of a bar, sufficiently strong to produce a ring instability in the disc of the bullet (e.g., Buta \&{} Combes 1996; Tiret \&{} Combes 2007). We evolved the companion in isolation till the formation of the resonant ring and we adopt this configuration as initial condition. This choice was made to test whether the companion of Arp~147 is a resonant ring galaxy. The resonant ring of the bullet survives relatively unperturbed for the first $\lesssim{}50$ Myr after the interaction, but it is more and more perturbed in the later stages of the evolution. This result cannot be used to constrain the properties of the companion of Arp~147, since we consider only one possible model for the companion. However, this may indicate either that the resonant ring of Arp~147 companion is more stable than our model against the perturbation (e.g., because of a more massive halo), or that we observe the Arp~147 system $\lesssim{}50$ Myr after the collision. Both cases are allowed by the data (e.g., Fogarty et al. 2011).  A further possibility is that even the ring of the companion has a collisional origin (that is, it was formed in the same interaction as the ring of the target). This hypothesis was proposed, e.g., for II~Hz~4 (Lynds \&{} Toomre 1976). Although we cannot exclude this scenario for Arp~147, there are various hints against it. First, the primary is an empty ring, produced by a collision with a large impact parameter, whereas the secondary is a symmetric ring, with a centred nucleus (although the system is observed with a large inclination), consistent with a nearly axisymmetric collision. These two morphologies are unlikely to form as a consequence of the same interaction. Second, the ring of the secondary is very smooth and regular, as observed in resonant ring galaxies (e.g., NGC~3081, Buta \&{} Purcell 1998).

\begin{table*}
\begin{center}
\caption{Differences in the initial conditions between runs.}
 \leavevmode
\begin{tabular}[!h]{llllllllll}
\hline
& & Run A
 & Run B
 & Run C
 & Run D
 & Run E
 & Run F
 & Run G
 & Run H\\
\hline
Target  & DM Mass [$10^{11}$ M$_\odot{}$]        & 7.0   & 7.0  & 7.0  & 7.0  & 7.0  & 7.0  & 7.0 & 2.1\\
        & M$_\ast{}$ [$10^{10}$ M$_\odot{}$]      & 5.66  & 7.06 & 7.77 & 8.13 & 8.13 & 8.13 & 8.13 & 8.13\\
       & Gas Mass [$10^{10}$ M$_\odot{}$]          & 2.82  & 1.42 & 0.71 & 0.35 & 0.35 & 0.35 & 0.35 & 0.35\\
\noalign{\vspace{0.1cm}}
Bullet & DM Mass   [$10^{11}$ M$_\odot{}$]         & 3.7 & 3.7 & 3.7 & 3.7 & 3.7 & 7.4 & 7.4 & 2.2 \\
\noalign{\vspace{0.1cm}}
\multicolumn{2}{l}{Impact Parameter $b$ [kpc]}         & 8   & 8   & 8   & 8  & 12   & 8  & 12 & 8\\
\multicolumn{2}{l}{Inclination $\alpha{}$ [$^\circ$]}         & 7   & 7   & 7   & 7  & 7   & 7  & 7 & 0 \\
\noalign{\vspace{0.1cm}}
\hline
\end{tabular}
\footnotesize{\\In this Table, we show only the parameters that have been changed in different runs.}
\end{center}
\end{table*}

\begin{figure}
\center{{
\epsfig{figure=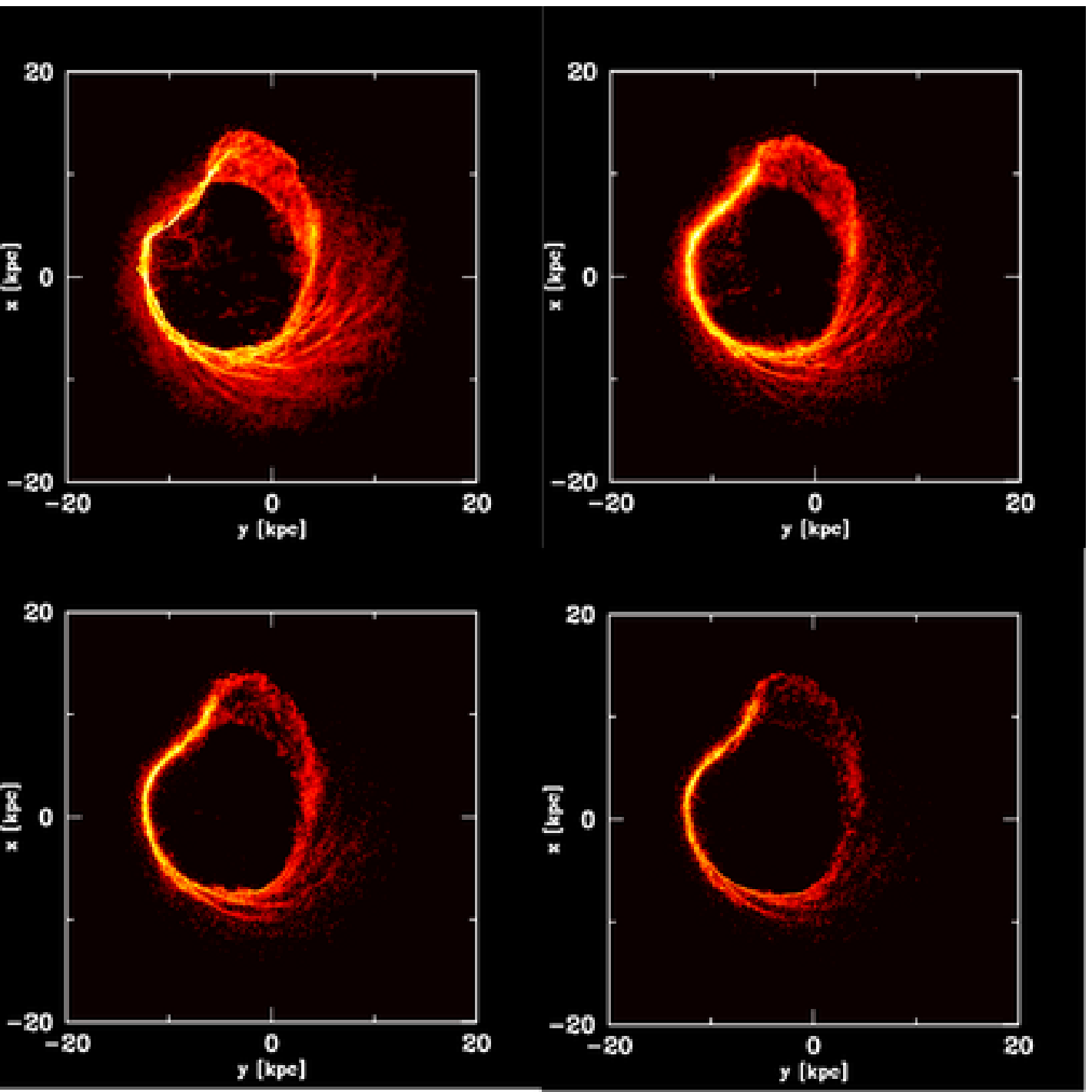,width=8.5cm} 
}}
\caption{\label{fig:fig2}
Mass density of gas of the target galaxy, projected in the $x-y$ plane,  at $t=50$ Myr after the collision.  From left to right and from top to bottom: run A, B, C and D. The scale is logarithmic, ranging from $2.22$ to $1.40\times{}10^3$ M$_\odot{}$ pc$^{-2}$.
}
\end{figure}

\begin{figure}
\center{{
\epsfig{figure=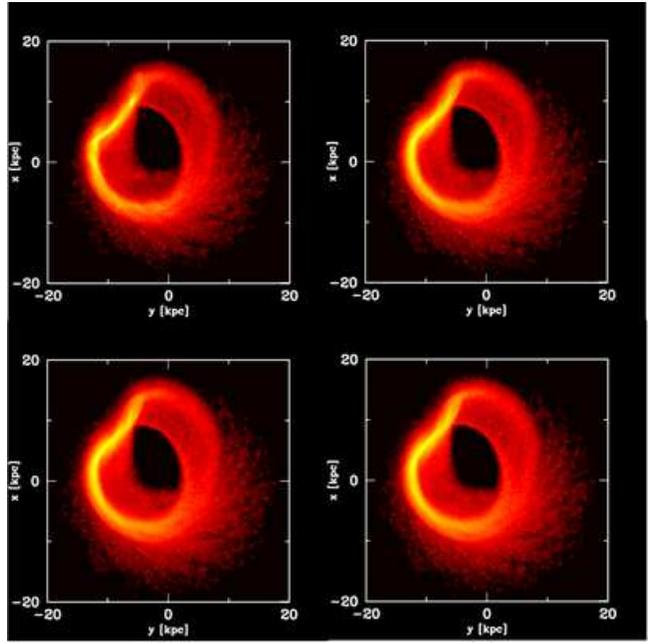,width=8.5cm} 
}}
\caption{\label{fig:fig3}
Mass density of stars of the target galaxy, projected in the $x-y$ plane,  at $t=50$ Myr after the collision. From left to right and from top to bottom: run A, B, C and D. The scale is logarithmic, ranging from $2.22$ to $2.22\times{}10^4$ M$_\odot{}$ pc$^{-2}$.
}
\end{figure}
Figs.~\ref{fig:fig2} and \ref{fig:fig3} show the projected density of gas and stars of the target (seen face-on), respectively, in runs A, B, C and D. The gas has a clumpy structure and is confined in a thinner ring than the stellar component. Both gas and stars form complete rings. In the simulations with a higher fraction of gas (runs A and B), the distribution of stars is clumpier, because the young stellar population is significant, and tracks the distribution of gas.
Instead, in  the runs with a lower fraction of gas (runs C and D), the old stars dominate the mass distribution.

\begin{figure}
\center{{
\epsfig{figure=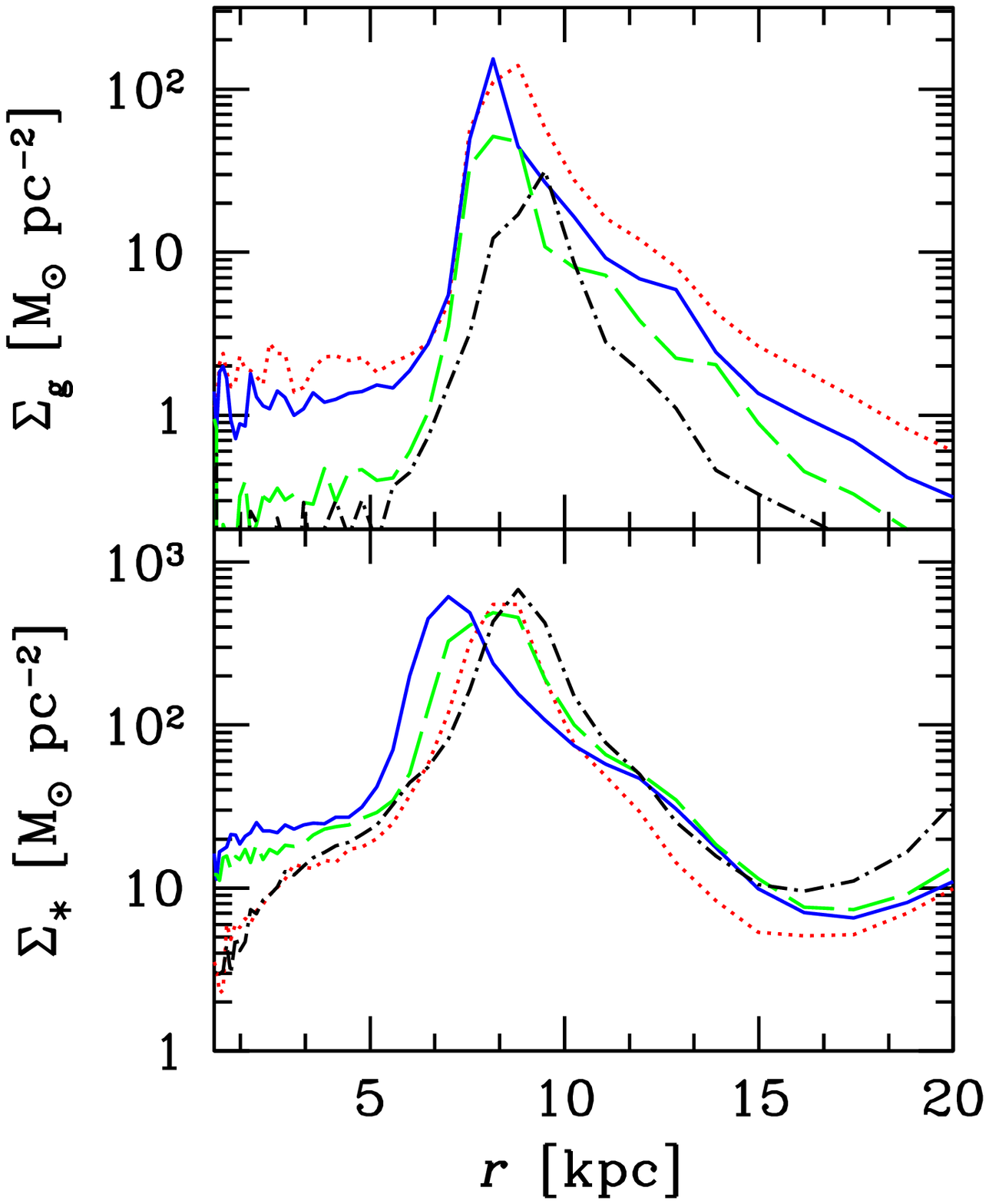,width=8.5cm} 
}}
\caption{\label{fig:fig4}
Top panel: surface density profile of gas of the target galaxy in run~A (dotted line, red on the web), run~B (solid line, blue on the web), run~C (dashed line, green on the web) and run~D (black dot-dashed line), at $t=50$ Myr after the collision.
Bottom panel: surface density profile of stars of the target galaxy, at $t=50$ Myr after the collision. Lines are the same as in the top panel.
}
\end{figure}
Fig.~\ref{fig:fig4} shows the surface density profile of gas (top) and stars (bottom) in runs A, B, C and D (which differ only for the gas to star mass fraction).  The four runs behave in a very similar way, with a strong density peak in correspondence of the ring (radius $r\sim{}6-10$ kpc, at $t=50$ Myr after the collision).

\begin{figure}
\center{{
\epsfig{figure=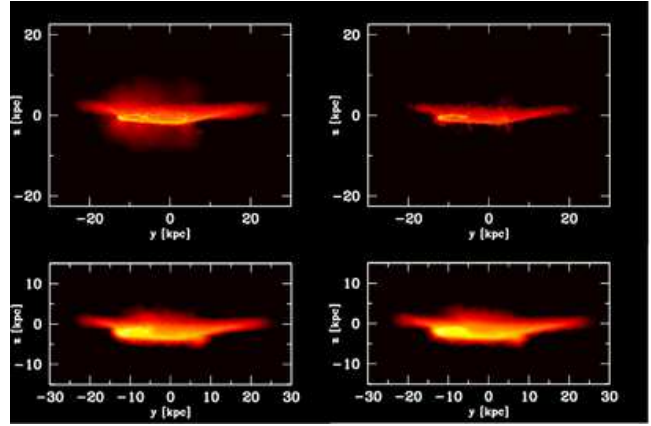,width=8.5cm} 
}}
\caption{\label{fig:fig5}
Top panels: mass density of gas of the target galaxy, projected in the $y-z$ plane,  at $t=50$ Myr after the collision. From left to right: run A and D. 
Bottom panels: mass density of stars of the target galaxy, projected in the $y-z$ plane,  at $t=50$ Myr after the collision. From left to right: run A and D. The scale is logarithmic, ranging from $7.04\times{}10^{-2}$ to $7.04\times{}10^4$ M$_\odot{}$ pc$^{-2}$.
}
\end{figure}

Fig.~\ref{fig:fig5} shows the projected density of gas (top) and stars (bottom)
when the target galaxy is seen edge-on. Only runs~A and D are shown, but runs~B and C are intermediate cases between the plotted ones. We find that, in all the considered runs, the nucleus of the target galaxy is still inside the disc even in the ring phase: the collision has off-set the nucleus from the dynamical centre and buried it in the ring, but the nucleus is not vertically displaced out of the ring. Actually, the entire ring, including the nucleus, is vertically displaced (by $\lesssim{}4$ kpc) with respect to the initial plane of the target disc.

\subsection{Morphology of the ring in runs E, F, G and H: comparison with Gerber et al. (1992)}
\begin{figure}
\center{{
\epsfig{figure=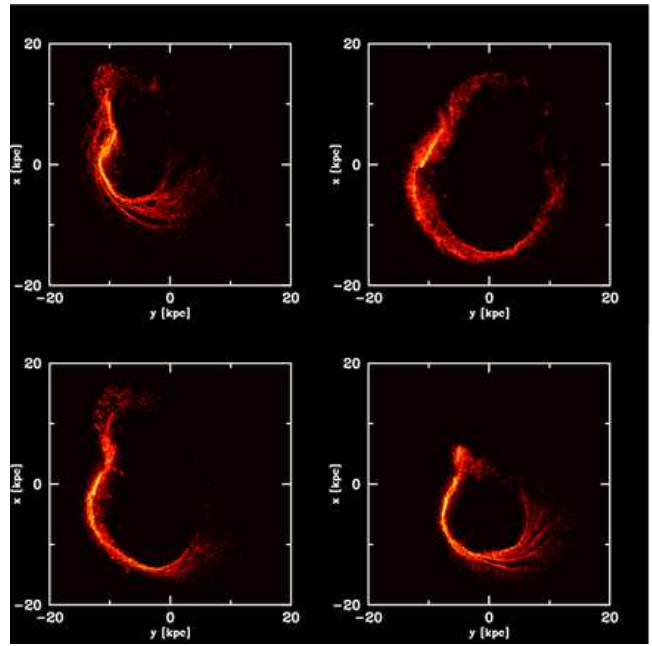,width=8.5cm} 
}}
\caption{\label{fig:fig6}
Mass density of gas of the target galaxy, projected in the $x-y$ plane,  at $t=50$ Myr after the collision.  From left to right and from top to bottom: run E, F, G and H. The scale is logarithmic, ranging from $2.22$ to $1.40\times{}10^3$ M$_\odot{}$ pc$^{-2}$.
}
\end{figure}
In our runs A, B, C and D, both the stellar and the gas ring are complete. This is not the case of  Gerber et al. (1992) runs: they conclude from their simulations that Arp~147 ring is incomplete and appears complete only because of projection effects. This discrepancy is likely due to differences in both the initial conditions and the resolution between our simulations and that by Gerber et al. (1992).  In particular, the main differences  between our runs A, B, C and D and those by Gerber et al. (1992) are the following. (i) Gerber et al. (1992) assume a mass ratio $M_{\rm targ}/M_{\rm int}=$1 between target and intruder, respectively, while in the runs shown above we assume $M_{\rm targ}/M_{\rm int}=$2. (ii) There is no DM halo in the galaxies simulated by  Gerber et al. (1992), but just a spherical stellar halo (modelled as a King profile), whereas we simulate a NFW DM halo. Furthermore, the halo to disc mass ratio in Gerber et al. (1992) is only $\sim{}2.5$. (iii) The mass and spatial resolution of  Gerber et al. (1992) are a factor of $\ge{}10$ worse than those of our simulations. (iv) In our runs~A, B, C and D, the impact parameter is $b=8$ kpc, corresponding to a distance $D=4-5$ kpc between the centres of mass of the two galaxies during the collision. In the paper by Gerber et al. (1992), $D=7$ kpc, which means that the collision is much more off-centre with respect to our simulations. (v) The simulations by Gerber et al. (1992) have no inclination angle.

To check the importance of these differences, we perform runs~E, F, G and H, exploring various parameters. In particular, runs~E and G have a larger impact parameter ($b=12$ kpc), runs~F, G and H have $M_{\rm targ}/M_{\rm int}=$1. Run~H was set to reproduce most of the features of Gerber et al. (1992) simulations (i.e., a DM to baryon fraction $\sim{}2.5$,  $M_{\rm targ}/M_{\rm int}=$1, no inclination angle,  $D=7$ kpc).

Fig.~\ref{fig:fig6} shows the density map of gas (face-on) in simulations E, F, G and H, at $t=50$ Myr after the interaction. The ring is incomplete in these four runs, especially in run~G (with $M_{\rm targ}/M_{\rm int}=$1 and $b=12$ kpc, bottom left-hand panel). In runs E and G (top and bottom left-hand panel) the partial ring is also strongly deformed, as a consequence of the large impact parameter, combined with the non-zero inclination angle (which induces some degree of warping). 

The ring in run~H, although incomplete, is much more circular and less deformed than the other rings: this is due to the absence of warping in runs with zero inclination angle. The fact that the halo-to-disc mass ratio is smaller in run~H than in the other runs has no significant effect on the propagation of the ring, at least in the initial stages. This is a consequence of the fact that the formation of the ring occurs in the inner regions of the target, that are baryon-dominated (we remind that the baryon mass in run~H is the same as in the other runs). Finally, the ring in run~H is smaller (by a factor of $\sim{}1.5$) than the ring in run~F, at the same time. This is likely due to the fact that the mass of the intruder is smaller (by a factor of $\sim{}3$) in run~H with respect to run~F\footnote{The total mass of the target is also smaller by a factor of $\sim{}3$ in run~H with respect to run~F, but the mass of the target in the inner regions is substantially unchanged, as the baryon mass is unchanged.}. In fact, we know from the analytic model that the radial velocity perturbation, in the impulse approximation, scales with the mass of the intruder (see, e.g., equation 2 of Struck-Marcell \&{} Lotan 1990).

In summary, a larger impact parameter is crucial to produce incomplete rather than complete rings, in combination with other parameters (e.g., the target-to-intruder mass ratio). Therefore, we suggest that RE galaxies form partly as complete and partly as incomplete rings, depending on the impact parameter, on the involved galaxy masses, potentials and scale-lengths.

From the analysis of runs A, B, C and D, we concluded that the nucleus of the target galaxy is not vertically displaced out of the ring (the entire ring, including the nucleus, is vertically displaced by $\lesssim{}4$ kpc with respect to the initial plane of the target disc, Fig.~\ref{fig:fig5}). Instead, Gerber et al. (1992) propose that the nucleus is displaced out of the plane defined by the ring by $\approx{}3$ kpc and it appears buried in the ring because of projection effects. 

\begin{figure}
\center{{
\epsfig{figure=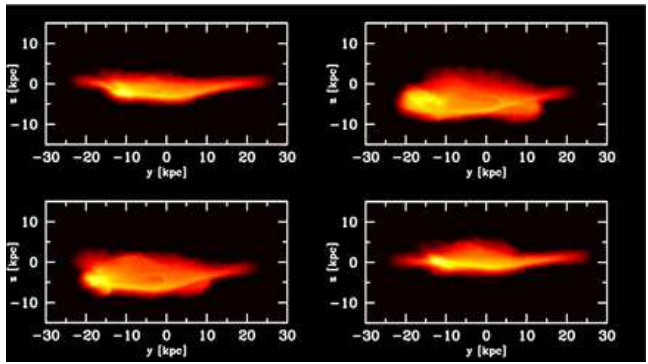,width=8.5cm} 
}}
\caption{\label{fig:fig7}
Mass density of stars of the target galaxy, projected in the $y-z$ plane,  at $t=50$ Myr after the collision. From left to right and from top to bottom: run E, F, G and H. The scale is logarithmic, ranging from $7.04\times{}10^{-2}$ to $7.04\times{}10^4$ M$_\odot{}$ pc$^{-2}$.
}
\end{figure}
Fig.~\ref{fig:fig7} shows the edge-on projected density of stars in runs E, F, G and H. The edge-on view of the ring in these runs is more perturbed than in our fiducial runs (A, B, C and D). In particular, the disc is much thicker in those simulations where $M_{\rm targ}/M_{\rm int}=$1 (runs F, G and H). However, we cannot conclude that the nucleus is completely displaced with respect to the rest of the ring. We cannot make more accurate comparisons, as Gerber et al. (1992) do not provide more information about the displacement they find. We expect that the possible differences between our simulations and those by Gerber et al. (1992) are connected with the different resolution or with the stability of the initial conditions.



As we discussed in Section~2, we decided to simulate the target as a bulgeless galaxy. However, we also run a check case in which the target has a bulge and all the other properties are the same as in run~D (see Appendix~B). In particular, we want to check whether the nucleus of the target can be more easily displaced from the plane of the ring after the collision, if the target has a bulge. The results indicate that the presence of a bulge (instead of a nucleus) in the target does not change the  features of the interaction (e.g., the completeness of the ring, the position of the nucleus, the kinematics and SFR, etc.). In particular, there is no displacement of the nucleus from the plane of the ring after the collision.

\subsection{Kinematics}
\begin{figure*}
\center{{
\epsfig{figure=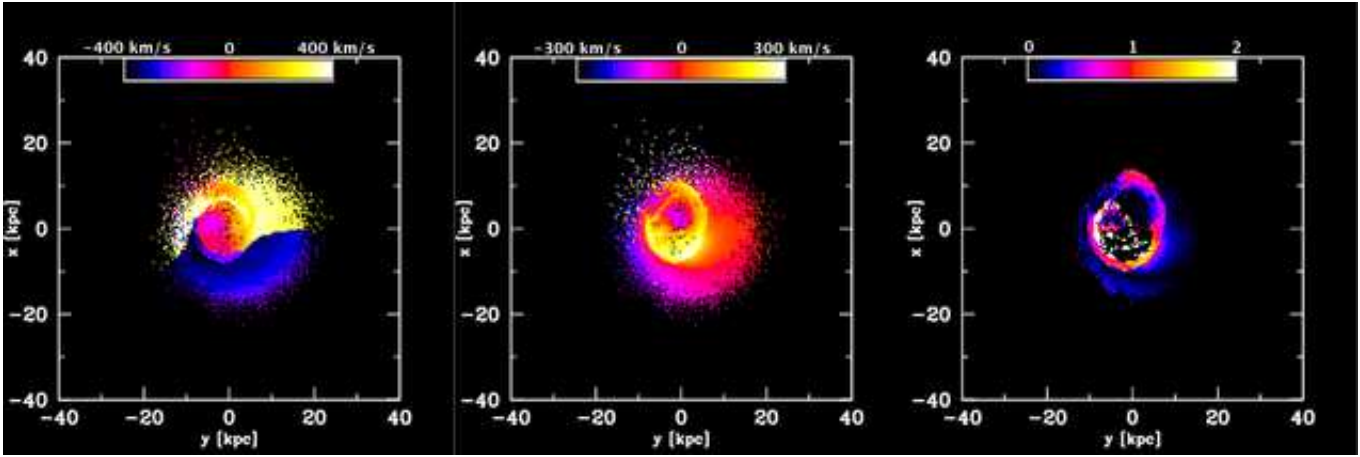,width=18cm} 
}}
\caption{\label{fig:fig8}
Velocity field of gas of the target galaxy in run~A,  at $t=50$ Myr after the collision. Left-hand panel: tangential velocity of gas (associated with the rotation of the disc) with respect to the disk of the target galaxy, projected in the $x-y$ plane. Central panel: radial velocity of gas (associated with the expansion or infall of the ring) with respect to the disk of the target galaxy, projected in the $x-y$ plane. Right-hand panel: ratio between the modulus of the radial velocity and the modulus of the tangential velocity, projected in the $x-y$ plane. The scale is linear and color coding is indicated in the panels.
}
\end{figure*}

The simulations also provide information about the kinematics of the interaction. Fig.~\ref{fig:fig8} shows the most relevant features of the velocity field of the simulated RE galaxy at $t=50$ Myr after the collision (run~A). The left-hand panel shows the tangential velocity $v_{\rm tan}$, i.e. the two-dimensional tangential velocity with respect to the disc of the target. Therefore, $v_{\rm tan}$ represents the rotational component of the target disc. The target galaxy rotates almost regularly for radii $r\gtrsim{}10$ kpc, whereas the rotation is strongly perturbed for $r<10$ kpc, because of the expanding ring. 

The central panel of Fig.~\ref{fig:fig8} shows the radial velocity $v_{\rm rad}$ with respect to the disc of the target, and measures the expansion of the ring. $v_{\rm rad}$ has various interesting features. First, the plot of $v_{\rm rad}$ highlights the high-velocity expanding ring at $r\approx{}10$ kpc. The ring has a maximum of the expansion velocity for the negative $x-$axis (which corresponds to the southern part of Arp~147). 

Second, the matter in the outer parts of the disc ($r>10$ kpc) in the region of the negative $y-$axis (which corresponds to the eastern part of Arp~147) has negative values of $v_{\rm rad}$ and  is falling toward the centre. This means that the densest part of the ring (which corresponds to the negative $y-$axis, see Figs.~\ref{fig:fig2} and \ref{fig:fig3}) is composed of both inner particles that are directed outwards and outer particles that are directed inwards. This result agrees with the predictions of the analytic caustic theory (see, e.g., Struck-Marcell \&{} Lotan 1990; Appleton \&{} Struck-Marcell 1996; Struck 2010).

Third, the matter in the outer parts of the disc ($r>10$ kpc) in the region of\
 the positive $y-$axis has $v_{\rm rad}\sim{}0$: it appears still unperturbed by the collision. Actually, this part of the target galaxy is the farthest from the position of the impact with the intruder.
The right-hand panel of Fig.~\ref{fig:fig8} shows the ratio between $|v_{\rm rad}|$ and $|v_{\rm tan}|$ and highlights the strong expansion of the ring, that dominates locally over the rotation.

Table~3 shows the average values of $|v_{\rm tan}|$ and $|v_{\rm rad}|$ in the inner part of the target galaxy ($r\le{}10$ kpc). These values are very similar for all the runs, as the orbits and the involved masses are almost the same. In particular, $\langle{}|v_{\rm rad}|\rangle{}\sim{}140\pm{}80$ km s$^{-1}$ and  $\langle{}|v_{\rm tan}|\rangle{}\sim{}200\pm{}110$ km s$^{-1}$ (with the marginal exception of runs G and H). The radial velocity $\langle{}|v_{\rm rad}|\rangle{}$ is comparable to the value of the expansion velocity derived by Fogarty et al. (2011) for Arp~147 ($V_{\rm exp}=113\pm{}4$ km s$^{-1}$), whereas $\langle{}|v_{\rm tan}|\rangle{}$ is quite higher than the observed rotational velocity ($V_{\rm rot}=47\pm{}8$ km s$^{-1}$). However, the value of $\langle{}|v_{\rm tan}|\rangle{}$ obtained from the simulations cannot be directly compared with the observations of $V_{\rm rot}$, because these two quantities are defined and derived in substantially different ways. We also stress that the relative velocity between the simulated galaxies is quite different from the value obtained by Fogarty et al. (2011). Furthermore, the uncertainty associated with $\langle{}|v_{\rm tan}|\rangle{}$ and with $\langle{}|v_{\rm rad}|\rangle{}$ is so large (Table~3) that these are consistent with the observational results by Fogarty et al. (2011) for Arp~147.
\begin{table}
\begin{center}
\caption{Kinematics at $t=50$ Myr after the interaction.}
 \leavevmode
\begin{tabular}[!h]{lll}
\hline
 & $\langle{}|v_{\rm rad}|\rangle{}$ [km s$^{-1}$]
 & $\langle{}|v_{\rm tan}|\rangle{}$[km s$^{-1}$]\\
\hline
Run A & $134\pm{}82$  & $186\pm{}117$ \\
Run B & $135\pm{}79$  & $187\pm{}115$ \\
Run C & $144\pm{}82$  & $207\pm{}115$ \\
Run D & $149\pm{}86$  & $218\pm{}115$ \\
Run E & $140\pm{}104$ & $216\pm{}114$ \\
Run F & $143\pm{}95$  & $156\pm{}128$ \\
Run G & $174\pm{}100$  & $217\pm{}130$ \\
Run H & $114\pm{}87$  & $193\pm{}103$ \\
\noalign{\vspace{0.1cm}}
\hline
\end{tabular}
\end{center}
\end{table}
\subsection{Star formation}
In the simulations, the SF and feedback recipes of Stinson et al. (2006) are adopted. Three parameters characterize the SF and feedback recipe: (a) the SF threshold $n_{\rm SF}$, (b) the SF efficiency $\epsilon_{\rm SF}$, and (c) the fraction of SN energy that couples to the interstellar medium (ISM) $\epsilon_{\rm SN}$.      
SF occurs when cold ($T<3\times 10^4$ K), virialized gas reaches a threshold density $n_{\rm SF}=5$ atoms cm$^{-3}$ 
and is part  of a converging flow. It proceeds at a rate 
\begin{equation}
\frac{{\rm d}\rho_*}{{\rm d}t}=\epsilon_{\rm SF}\,{} \frac{\rho_{\rm gas}}{t_{\rm dyn}} \propto \rho_{\rm gas}^{1.5}
\label{eq:KS}
\end{equation}
(i.e. locally enforcing a Schmidt law), where $\rho_*$ and $\rho_{\rm gas}$ are the stellar and gas densities, and $t_{\rm dyn}$ 
is the local dynamical time. We choose $\epsilon_{\rm SF}=0.1$.
The relatively high SF density threshold that we adopt is the same as that used in recent cosmological simulations that form realistic
galaxies (Guedes et al. 2011) and it has been shown to have a key role in determining a realistic inhomogeneous structure of the ISM (see, e.g., Mayer 2011). Its value is determined by enforcing that at least one SPH kernel
(32 particles) is contained within a spherical volume of diameter equal to the local Jeans length for gas at density $n_{\rm SF}$ at the
lowest temperatures allowed by the adopted cooling function (a few thousand K).
In the blastwave feedback model of Stinson et al. (2006), the gas heated by the SN energy is not allowed to cool for a timescale of the order of 10 Myr, the exact value being self-consistently determined by the blastwave solution of the McKee \& Ostriker (1977) model, that assumes a blastwave produced by the collective effect of many type II SNae. For  type I SNae, the energy is instead radiated
away on the cooling time (since they explode on longer timescales they will explode at significantly different time and would hardly produce a collective blastwave).

\begin{figure}
\center{{
\epsfig{figure=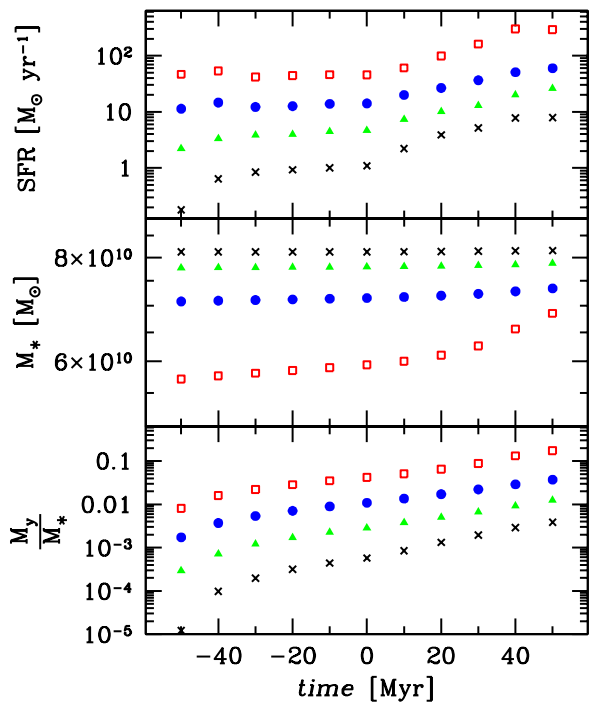,width=8cm} 
}}
\caption{\label{fig:fig9}
Top panel: SFR as a function of time in run~A (open squares, red on the web), run~B (filled circles, blue on the web), run~C (filled triangles, green on the web) and run~D (black crosses). 
Central panel: total mass in stars (M$_\ast{}$) of the target galaxy. Symbols are the same as in top panel. 
Bottom panel: total mass of young stars  (M$_{\rm y}$) normalized to the total mass in stars of the target galaxy. Symbols are the same as in top panel. 
Time $t=0$ corresponds to the collision. 
}
\end{figure}

To assess the effects of the SFR in our simulations, we will focus on runs A, B, C and D, that differ only for the gas-to-star fraction.
The top panel of Fig.~\ref{fig:fig9} shows the SFR as a function of time in the four considered simulations. In all the simulations, the SFR has approximately the same trend: it increases sensibly after the galaxy interaction. The main difference among the simulations is the normalization of the SFR, that spans from $\approx{}8$ M$_\odot{}$ yr$^{-1}$ in run~D to $\approx{}300$ M$_\odot{}$ yr$^{-1}$ in run~A. 
Observations of Arp~147 (Rappaport et al. 2010; Fogarty et al. 2011) indicate a SFR$\sim{}4-12$ M$_\odot{}$ yr$^{-1}$, consistent with runs~C and~D.

The central and bottom panels of Fig.~\ref{fig:fig9} show the time evolution of the total stellar mass of the target (M$_\ast{}$) and of the mass fraction of young stars (M$_{\rm y}/$M$_\ast{}$; we define as young stars those stars that have an age $t_{\rm age}\le{}100$ Myr at $t=50$ Myr after the interaction). The total stellar mass of the four simulations, at $t=50$ Myr after the interaction, spans from $7\times{}10^{10}$ to $8\times{}10^{10}$ M$_\odot{}$. Therefore, all these simulations are consistent with the stellar mass of Arp~147 (found to be M$_\ast{}=6.1-10.1\times{}10^{10}$ M$_\odot{}$ by Fogarty et al. 2011).

However, there are large differences in the role of the young stellar population between different runs. In fact, M$_{\rm y}/$M$_\ast{}$ ranges from $5\times{}10^{-3}$ (in run~D) to 0.17 (in run~A) at $t=50$ Myr. Since Fogarty et al. (2011) find that the mass fraction of young stars in Arp~147 is $\le{}1$ per cent, we confirm that only runs~C and~D are consistent with the observations of Arp~147.

 The difference in M$_{\rm y}/$M$_\ast{}$ among various runs is consistent with the difference expected from the Schmidt law. For example, there is a factor of $\sim{}34$ between run A and D. From the Schmidt law, we expect that the SFR scales with $\rho_{\rm gas}^{1.5}$, which means that there should be at least a factor of $\approx{}23$ difference between  M$_{\rm y}$ in runs A and D. Since the mass of old stars in run~D is a factor of $\sim{}1.4$ higher than in run~A, we expect a difference by a factor of $\sim{}32$ in M$_{\rm y}/$M$_\ast{}$  between  run~A and D.

\section{Conclusions}
In this paper, we investigate the formation of RE galaxies (i.e. of collisional ring galaxies with an empty ring) with N-body/SPH simulations. We have shown that RE galaxies can form via off-centre collisions (i.e. with a non-zero impact parameter), even for small inclination angles ($\sim{}0-7^\circ{}$).

In runs A, B, C and D (where $b=8$ kpc and $M_{\rm targ}/M_{\rm int}=$2), the ring is always complete and it is not produced by a projection effect. In runs with a larger impact parameter ($b=12$ kpc) or with $M_{\rm targ}/M_{\rm int}=$1, the ring looks incomplete, as proposed by Gerber et al. (1992). Therefore, we conclude that RE galaxies can have either complete or incomplete rings, depending on the parameters of the interaction. 

We find that the nucleus is displaced from the dynamical centre of the galaxy, as a consequence of the off-centre collision. However, the nucleus remains in the (perturbed) plane of the ring. The entire ring (together with the nucleus) is slightly ($\lesssim{}4$ kpc) vertically displaced with respect to the previous plane of the disc of the target galaxy. This result is  different from the findings by Gerber et al. (1992), who claim that the nucleus appears buried within the ring only as a consequence of projection effects. We expect that this discrepancy is related to the lower resolution of Gerber et al. (1992). 

The kinematics of the ring was studied in detail, confirming the predictions by the analytic caustic theory (e.g., Struck-Marcell \&{} Lotan 1990). We compare our simulations with the observations of Arp~147, finding similar expansion velocity in the ring. 

Finally, we study the SF history of the RE galaxy, finding that the encounter enhances the SFR of the target galaxy. Runs~C and~D match the observational properties of Arp~147 (Rappaport et al. 2010; Fogarty et al. 2011), having average SFR$\sim{}10-20$ M$_\odot{}$ yr$^{-1}$, total star mass M$_\ast{}=8\times{}10^{10}$ M$_\odot{}$ and  mass fraction of young stars M$_{\rm y}/{\rm M}_\ast{}\sim{}(0.5-1.6)\times{}10^{-2}$.

We stress that we adopt a recent implementation of the SFR and of SNae in our simulations, whose results are in good agreement with observations (see. e.g., Stinson et al. 2006, 2009). However, we do not model the chemical evolution of the gas, which may be very interesting in ring galaxies, as pointed out by recent papers (e.g., Bournaud et al. 2007; Mapelli et al. 2009, 2010; Michel-Dansac et al. 2010).
Future N-body models of RE galaxies (and, in general, of collisional ring galaxies) should account for multi-phase gas (to describe the evolution of molecular and atomic gas) and include metallicity evolution, to shed light on the SF history of ring galaxies.






\section*{Acknowledgments}
We thank the referee, C. Horellou, for her comments that significantly improved the paper. We thank the authors of gasoline (especially J. Wadsley, T. Quinn and J. Stadel). We also thank L.~Widrow for providing us the code to generate the initial conditions. 
The simulations were performed with the {\it lagrange} cluster at the Consorzio Interuniversitario Lombardo per L'Elaborazione Automatica (CILEA) and with the {\it Yoda} linux cluster at the University of Insubria. We thank S. Rappaport, E. Ripamonti, A. Wolter, D. Fiacconi and A. A. Trani for useful discussions.

\appendix
\section{Large inclination angle}
\begin{figure}
\center{{
\epsfig{figure=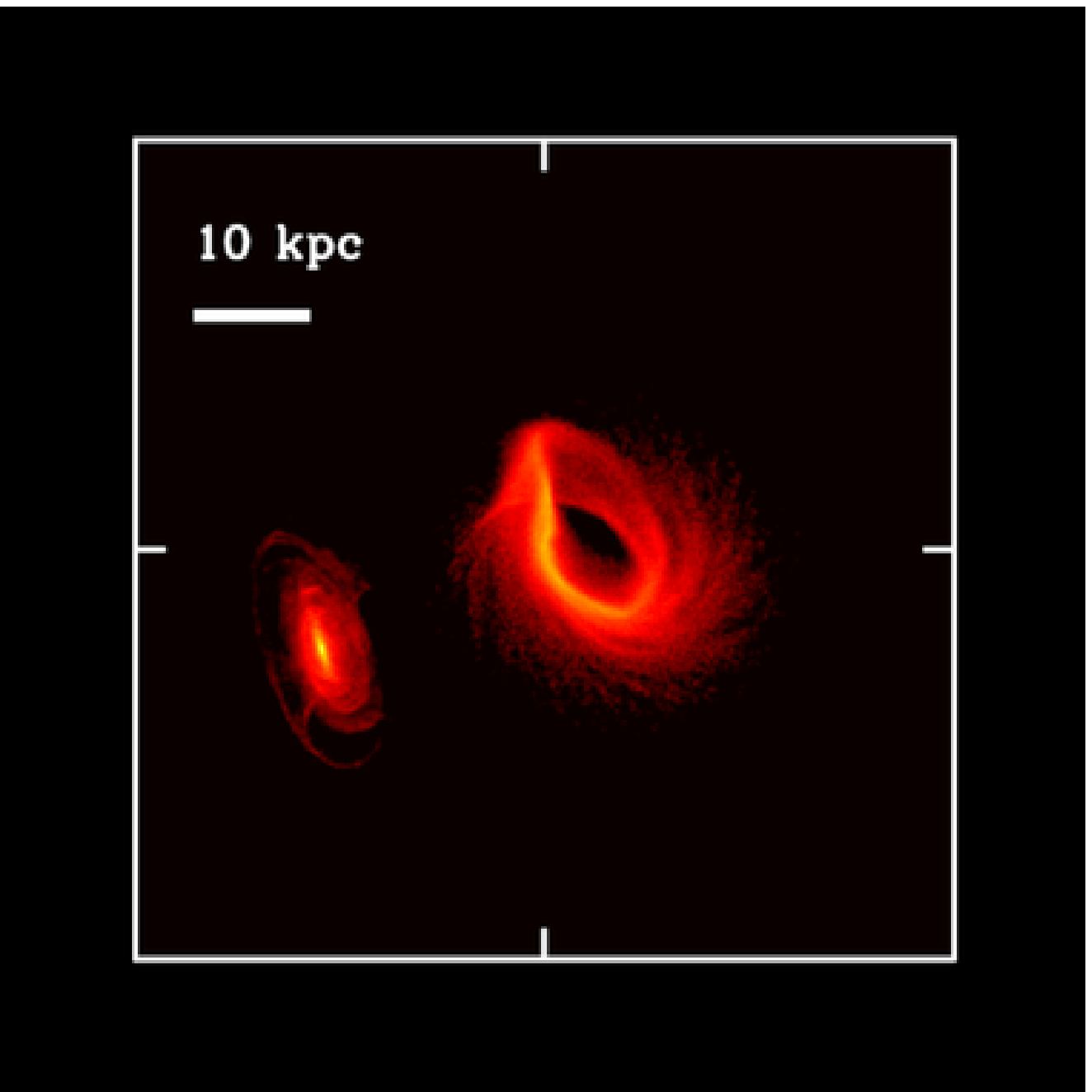,width=8.5cm} 
}}
\caption{\label{fig:figA1}
Projected mass density of stars and gas in the run with $\alpha{}=33^\circ$. The simulated target galaxy has been rotated by 20$^{\circ}$ along the $x$-axis in the direction of the $y$-axis and by 20$^{\circ}$ counterclockwise, to match the observations of Arp~147. The scale is logarithmic, ranging from $11.15$ to $7.04\times{}10^4$ M$_\odot{}$ pc$^{-2}$. 
}
\end{figure}

In this appendix, we show a simulation where the inclination angle is $\alpha{}=33^\circ$ (in agreement with the estimates for Arp~147 by Fogarty et al. 2011). The velocity components were rearranged so that the impact parameter is still 8 kpc. All the other initial conditions of this simulation are the same as in run~D. From Fig.~\ref{fig:figA1}, it is evident that the ring of the simulated galaxy is strongly warped, because of the large inclination angle. We will investigate in detail the effects of warping in a forthcoming paper.

\section{Target galaxy with bulge}
\begin{figure}
\center{{
\epsfig{figure=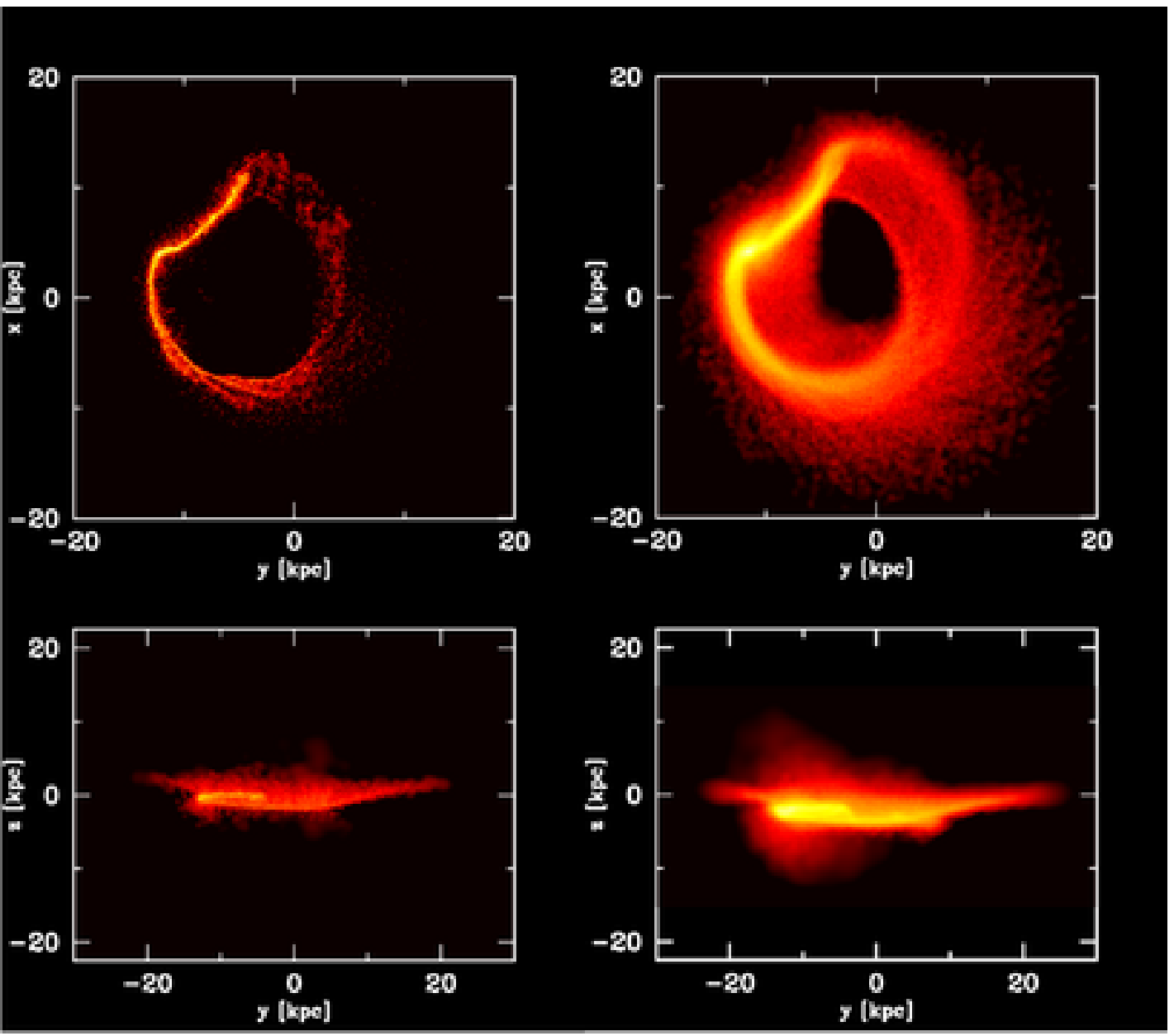,width=8.5cm} 
}}
\caption{\label{fig:figB1}
Mass density of gas and stars of the target galaxy in the check run with bulge,  at $t=50$ Myr after the collision.  Top panels: mass density of gas (left-hand panel) and of stars (right-hand panel) projected in the $x-y$ plane (i.e., face-on). The scale is logarithmic, ranging from $2.22$ to $1.40\times{}10^3$ M$_\odot{}$ pc$^{-2}$ and from 2.22 to $2.22\times{}10^4$ M$_\odot{}$ pc$^{-2}$ in the left-hand and in the right-hand panel, respectively.
Bottom panels: the same as the top panels, but the target galaxy is projected in the $y-z$ plane (i.e., edge-on). The scale is logarithmic, ranging from $7.04\times{}10^{-2}$ to $7.04\times{}10^4$ M$_\odot{}$ pc$^{-2}$. 
}
\end{figure}
In this appendix, we discuss the results of a simulation in which the target galaxy has a bulge (with a total mass of $1.2\times{}10^{10}$ M$_\odot{}$). The other properties of this simulation are the same as those of run~D. Fig.~\ref{fig:figB1} shows the projected density of the target galaxy face-on (top) and edge-on (bottom) for both the gaseous (left-hand) and the stellar component (right-hand panel). No significant differences can be observed with respect to run~D in Figs.~\ref{fig:fig2}, \ref{fig:fig3} and \ref{fig:fig5}. In the case of the target with bulge, the ring is still complete, and the nucleus/bulge is buried within the ring, but it is not vertically displaced. The SFR (6 M$_\odot{}$ yr$^{-1}$ at $t=50$ Myr after the collision) is also very similar to run~D (8 M$_\odot{}$ yr $^{-1}$). Furthermore, the kinematics of the ring cannot be distinguished from run~D, as in the simulation with bulge $v_{\rm rad}=149\pm{}89$ km s$^{-1}$ and $v_{\rm tan}=201\pm{}111$ km s$^{-1}$ (we remind that $v_{\rm rad}=149\pm{}86$ km s$^{-1}$ and $v_{\rm tan}=218\pm{}115$ km s$^{-1}$ in run~D, see Table~3). Therefore, the assumption about the bulge of the target does not significantly affect our results, in the case of very late-type galaxies, such as Arp~147.

\begin{thebibliography}{}

\bibitem{}Appleton P. N., James R. A., 1990, International Conference on Dynamics and Interactions of Galaxies, 200
\bibitem{}Appleton P. N., Struck-Marcell C., 1987a, ApJ, 312, 566
\bibitem{}Appleton P. N., Struck-Marcell C., 1987b, ApJ, 318, 103
\bibitem{}Appleton P. N., Struck-Marcell C., 1996, {\it Fundamentals of Cosmic Physics}, 16, 111
\bibitem{}Bell E. F., McIntosh D. H., Katz N., Weinberg M. D., 2003, ApJS, 149, 289
\bibitem{}Bournaud F., Duc P.-A., Brinks E., Boquien M., Amram Ph., Lisenfeld U., Koribalski B. S., Walter F., Charmandaris V., 2007, Science, 316, 1166
\bibitem{}Buta R., Combes F., 1996, Fundamentals of Cosmic Physics, 17, 95
\bibitem{}Buta R., Purcell G. B., 1998, AJ, 115, 484
\bibitem{}de Vaucouleurs G., de Vaucouleurs A., Corwin H. G. Jr., Buta R. J., Paturel G., Fouque P., 1991, Third Reference Catalogue of Bright Galaxies, Springer-Verlag Berlin Heidelberg New York

\bibitem{}Few J. M. A., Madore B. F., 1986, MNRAS, 222, 673
\bibitem{}Fogarty, L., Thatte N., Tecza M., Clarke F., Goodsall T., Houghton R., Salter G., Davies R., Kassin S., 2011, MNRAS, 417, 835
\bibitem{}Fosbury R. A. E., Hawarden T. G., 1977, MNRAS, 178, 473
\bibitem{}Geha M., Blanton M. R., Masjedi M., West A. A., 2006, ApJ, 653, 240 
\bibitem{}Gerber R. A., Lamb S. A., Balsara D. S., 1992, ApJ, 399, L51
\bibitem{}Gerber R. A., Lamb S. A., Balsara D. S., 1996, MNRAS, 278, 345
\bibitem{}Ghosh K. K., Mapelli M., 2008, MNRAS, 386, L38
\bibitem{}Governato F., Brook C., Mayer L., Brooks A., Rhee G., Wadsley J., Jonsson P., Willman B., Stinson G., Quinn T., Madau P., 2010, Nature, 463, 203
\bibitem{}Guedes J., Callegari S., Madau P., Mayer L., 2011, ApJ, 742, 76
\bibitem{}Haynes M. P., Giovanelli R., Chamaraux P., da Costa L. N., Freudling W., Salzer J. J., Wegner G., 1999, AJ, 117, 2039
\bibitem{}Hernquist L., 1993, ApJS, 86, 389
\bibitem{}Hernquist L., Weil M. L., 1993, MNRAS, 261, 804
\bibitem{}Higdon J. L., 1995, ApJ, 455, 524
\bibitem{}Higdon J. L., 1996, ApJ, 467, 241
\bibitem{}Horellou C., Combes F., 2001, Ap\&{}SS, 276, 1141
\bibitem{}Katz N., 1992, ApJ, 391, 502
\bibitem{}Kuijken K., Dubinski J., 1995, MNRAS, 277, 1341
\bibitem{}Lynds R., Toomre A., 1976, ApJ, 209, 382
\bibitem{}Mapelli M., Moore B., Giordano L., Mayer L., Colpi M., Ripamonti E., Callegari S., 2008a, MNRAS, 383, 230
\bibitem{}Mapelli M., Moore B., Ripamonti E., Mayer L., Colpi M., Giordano L., 2008b, MNRAS, 383, 1223
\bibitem{}Mapelli M., Colpi M., Zampieri L., 2009, MNRAS, 395, L71
\bibitem{}Mapelli M., Ripamonti E., Zampieri L., Colpi M., Bressan A., 2010, MNRAS, 408, 234
\bibitem{}Michel-Dansac L., Duc P.-A., Bournaud F., Cuillandre J.-Ch., Emsellem E., Oosterloo T., Morganti R., Serra P., Ibata R., 2010ApJ, 717, L143
\bibitem{}Marston A. P., Appleton P. N., 1995, AJ, 109, 1002
\bibitem{}Mayer L., 2011, EAS Publications Series, 48, 369
\bibitem{}Mayya Y. D., Bizyaev D., Romano R., Garcia-Barreto J. A., Vorobyov E. I., 2005, ApJ, 620L, 35
\bibitem{}Mihos J. C., Hernquist L., 1994, ApJ, 437, 611
\bibitem{}Navarro J. F., Frenk C. S., White S. D. M., 1996, ApJ, 462, 563 (NFW)
\bibitem{}Pilyugin L. S., V\'ilchez J. M., Thuan T. X., 2010, ApJ, 720, 1738 
\bibitem{}Rappaport S., Levine A., Pooley D., Steinhorn B., 2010, ApJ, 721, 1348
\bibitem{}Romano R., Mayya Y. D., Vorobyov E. I., 2008, AJ, 136, 1259
\bibitem{}Schultz A. B., Spight L. D., Rodrigue M., Colegrave P. T., Disanti M. A., 1991, Bulletin of the American Astronomical Society, 23, 953
\bibitem{}Smith B. J., Struck C., Nowak M. A., 2005, AJ, 129, 1350
\bibitem{}Stinson G., Seth A., Katz N., Wadsley J., Governato F., Quinn T., 2006, 373, 1074
\bibitem{}Stinson G., Dalcanton J. J., Quinn T., Gogarten S. M., Kaufmann T., Wadsley J., 2009, 395, 1455
\bibitem{}Struck-Marcell C., Lotan P., 1990, ApJ, 358, 99
\bibitem{}Struck C., 1997, ApJS, 113, 269
\bibitem{}Struck C., 2010, MNRAS, 403, 1516
\bibitem{}Theys J. C., Spiegel E. A., 1976, ApJ, 208, 650
\bibitem{}Thompson L. A., Theys J. C., 1978, ApJ, 224, 796
\bibitem{}Tiret O., Combes F., 2007, A\&{}A, 464, 517
\bibitem{}Toomre A., 1978, In: The large scale structure of the universe; Proceedings of the Symposium, Tallin, Estonian SSR, September 12-16, 1977. Dordrecht, D. Reidel Publishing Co., p. 109-116
\bibitem{}Wadsley J. W., Stadel J., Quinn T., 2004, New Astronomy, 9, 137
\bibitem{}Weilbacher P. M., Duc P.-A., Fritze-v. Alvensleben U., 2003, A\&{}A, 397, 545
\bibitem{}Widrow L. M., Dubinski J., 2005, ApJ, 631, 838
\bibitem{}Widrow L. M., Pym B., Dubinski J., 2008, ApJ, 679, 1239

\end{thebibliography}
\end{document}